\documentclass[aps,prl,preprint,showpacs]{revtex4}
\usepackage{amsmath}
\newcommand{\be}{\begin{equation}}
\newcommand{\ee}{\end{equation}}
\newcommand{\bea}{\begin{eqnarray}}
\newcommand{\eea}{\end{eqnarray}}

\usepackage{graphicx}
\begin{document}

\title{On the Microcanonical Entropy of a Black Hole} 

\author{Rajat K. Bhaduri}
\author{Muoi N. Tran}
\affiliation{Department of Physics and Astronomy, McMaster University,
Hamilton, Ontario, Canada L8S~4M1}
\author{Saurya Das} 
\affiliation{Department of Physics, University of Lethbridge, 
Lethbridge, Alberta, Canada T1K 3M4}

\date{\today}

\begin{abstract}
It has been suggested recently that the microcanonical entropy of a system  
may be accurately reproduced by including a logarithmic  
correction to the canonical entropy. In this paper we test this claim both 
analytically and numerically by considering three simple thermodynamic models 
whose energy spectrum may be defined in terms of one quantum number only, as 
in a non-rotating black hole. The first two pertain to collections of 
noninteracting bosons, with logarithmic and power-law spectra.
The last is an area ensemble
for a black hole with equi-spaced area spectrum. In this case, the many-body 
degeneracy factor can be obtained analytically in a closed form. 
We also show that in this model, the leading term in the entropy is proportional to the horizon area $A$, and the next term is $\ln A$ with a negative coefficient.
\end{abstract}
\vspace{0.5cm}


\maketitle

\section{Introduction}
The entropy of a macroscopic black hole is known to be proportional to the 
area of its horizon~\cite{area}, in units of the Planck length squared. 
It has also been shown by several authors using a variety of approaches that
the leading order correction to this is proportional 
to the logarithm of the area \cite{log} (also see refs. in \cite{das}). 
Recently, a universal  
form for the (negative) coefficient of this logarithmic term has been obtained 
in ref.~\cite{chatter} by assuming a power-law dependence of the area on the 
mass of the (non-rotating) black hole. 
For an isolated black hole, it is of course appropriate to consider the 
microcanonical entropy. For a quantum system, the microcanonical entropy 
may be defined uniquely in terms of the degeneracy of the state at a given 
energy, and it has no fluctuation in energy. The many-body degeneracy factor, 
for any nontrivial system, however, is exceedingly difficult to calculate. 
For this reason, it is desirable to approximate 
the microcanonical entropy by the canonical entropy (the leading term), 
minus a logarithmic term due to fluctuations in the canonical 
ensemble-averaged energy from the equilibrium value.  
The main objective of this paper is to test this formula  
quantitatively in three solvable models, where the microcanonical entropy can 
also be calculated exactly. The first two of these are systems of 
noninteracting bosons, and {\it not} related to black holes.  
These many-body systems, however, have eigenenergies 
that depend on a single quantum number, similar to a non-rotating quantum 
black hole. The microcanonical entropy of 
such systems is calculated exactly, and then compared with the approximate 
(canonical) formula. When the logarithmic correction to the canonical entropy  
are included in the canonical expression, its agreement 
with the microcanonical entropy improves markedly.  Next, we consider a canonical area ensemble with equi-spaced spectrum and distinguishable area components as model for a non-rotating black hole.  The area of the surface of the event horizon plays a role analogous to the energy \cite{krasnov}.  This equi-spaced spectrum was first proposed in the early seventies (see ref.\cite{bekmukh}) 
and later confirmed by several authors using different techniques
\cite{equi}.  The model has been studied earlier by Alekseev {\it et~al.}~\cite{aps} using the grand canonical ensemble.  In this paper, we are able to calculate the exact microcanonical entropy for this spectrum in a closed form, and obtain the area-law for the entropy.  We also show that the next term in the entropy is proportional to $\ln A$ with a negative coefficient.  This is also verified using the canonical ensemble when the area fluctuation is subtracted out.

The origin of the black hole entropy, in theories of quantum gravity, is 
believed to arise from the microstates that are generated by a quantum 
mechanical operator such as area. 
In standard statistical mechanics, when many particles in the 
mean-field model are trapped in a potential well, the microcanonical entropy 
of the many-body system at a (quantised) energy $E_n$ 
is obtained by taking the logarithm of the number of distinct microstates 
that all give the same energy $E_n$. To be more explicit, 
$E_n=\sum_{\{N_i\}_n}N_i\epsilon_i$, where $N_i$ is the number of particles 
with single-particle energy $\epsilon_i$. The set $\{N_i\}_n$ denotes a 
given occupancy configuration 
of single-particle levels that make up a microstate with total energy $E_n$. 
There may be $\Omega(E_n)$ such distinct microstates for an energy $E_n$, 
each denoted by a set $ \{N_i\}$, and the microcanonical entropy is then  
uniqely defined as
\be
{\cal S}(E_n) = k_B \ln(\Omega(E_n)), 
\label{mike}
\ee
where the Boltzmann constant $k_B$ will henceforth be put to unity. 
Similarly, in models of quantum gravity, the (macroscopic) area eigenvalue 
$A_n$ is taken to be coming 
from the elementary components $a_i$, such that 
$A_n=\sum_{\{N_i\}_n} N_i a_i$, 
where $N_i$ is the number of elementary components with area $a_i$~\cite{aps}. 
Each microstate is specified by a distinct set $\{N_i\}$, and there may be 
$\Omega(A)$ such microstates for a given area $A$. The microcanonical 
entropy is then ${\cal S}(A_n)=\ln (\Omega (A_n))$. 

\section{Energy spectrum depending on one quantum number}

Consider a many-body quantum system with eigenenergies $E_n$ that are 
completely specified by a single quantum number $n$, 
\be
E_n=f(n),~~~~~~~~n=0,1,2,3,....
\label{night1}
\ee
where we assume $f(n)$ to be an arbitrary monotonous function with a 
differentiable inverse, $f^{-1}(x)=F(x)$, such that $n=F(E_n)$. The degeneracy 
of the states at energy $E_n$ is given by $\Omega(E_n)$, a function 
characterising the quantum spectrum. At this point, we need not assume that 
this many-body system is described in the mean-field picture. The quantum 
density of the system is defined as 
\be
\rho(E)=\sum_{n} \Omega(E_n) \delta (E-E_n)~.
\label{base}
\ee 
Using general properties of the delta function, we write 
\be
\delta(E-E_n)=\delta(E-f(n))=\delta(n-F(E))|F'(E)|~,
\ee
where the prime denotes differentiation with respect to the continuous 
variable $E$. The quantum density of states of the system is then given by 
\be
\rho(E)=\Omega(E)|F'(E)|\sum_{n=0}^{n=\infty} \delta(n-F(E))~.
\label{night2}
\ee 
We know from the Poisson summation formula 
\be
\sum_{n=-\infty}^{\infty} \delta(E-n)=\sum_{k=-\infty}^{\infty}\exp(2i\pi k).
\ee
So, if $E\geq 0$, we get 
\be
\sum_{n=0}^{\infty} \delta(E-n)=1+2\sum_{k=1}^{\infty} {\rm cos}(2\pi kE)~.
\ee 
We thus obtain~\cite{brack}, from Eq.~(\ref{night2}),
\be
\rho(E)=\Omega(E)|F'(E)|[1+2\sum_{k=1}^{\infty} {\rm cos}(2\pi kF(E))]~.
\label{night3}
\ee
We assume the function $\Omega(E)$ of the continuous variable $E$ to be 
smooth. The first term on the RHS of the above relation is then the  
smoothly varying part 
of the density of states, while the second part consists of  the oscillating 
components coming from the discreteness of the energy levels. For a 
macroscopic system with large $E$, the oscillating part may be neglected. 
We then obtain the important relation 
\be
\tilde{\rho}(E)=\Omega(E)|F'(E)|~,
\label{night4}
\ee
where $\tilde{\rho}(E)$ denotes the averaged smooth density of states. 
Now we specialize to a system of $N$ noninteracting particles (or in a mean 
field) constituting the many-body system. Then the degeneracy  
$\Omega(E_n)$ is just the number of distinct microstates that all have the   
same energy $E_n$, as described in the introduction. The microcanonical 
entropy is given by Eq.~(\ref{mike}), which, using Eq.~(\ref{night4}), may 
now be expressed as  
\be
{\cal S}\simeq \ln [\tilde{\rho}(E)|F'(E)|^{-1}]~,
\label{night5}
\ee
where the oscillating part has been dropped.

Our next task is to calculate $\tilde{\rho}(E)$ for a many-body system. 
This may be obtained by considering the canonical partition function of 
the $N$-particle system, and taking its inverse Laplace transform using the 
saddle-point method~\cite{das}. 
The canonical $N-$particle partition function is given by 
\be
Z(\beta) = \sum_{n}\Omega(E_n) \exp(-\beta E_n)~= 
\int_0^{\infty} \rho(E) \exp(-\beta E) dE~, 
\label{zce}
\ee 
where $\rho(E)$ is defined by Eq.~(\ref{base}). Note that we are using the 
canonical ensemble only as a tool to obtain $\tilde{\rho}(E)$ by Laplace 
inversion with respect to $\beta$, which is just a variable of integration 
along the imaginary axis. The saddle-point approximation simply gives the 
smooth part $\tilde{\rho}(E)$. The well-known result \cite{kubo} is    
\be
\tilde{\rho}(E) = \frac{\exp[S_C(\beta_0)]}{\sqrt{2\pi S_C''(\beta_0)}},
\label{rho2}
\ee 
where 
\be
S_C(\beta_0) = \beta_0 E + \log Z(\beta_0)
\label{entropy}
\ee
is the canonical entropy evaluated at the stationary point $\beta_0$, the 
prime denotes differentiation with respect to $\beta$.   
The energy $E$ is related to the saddle point $\beta_0$ via
\be
E=-\left( \frac{\partial \ln Z}{\partial \beta}\right)_{\beta_0}. 
\label{saddle}
\ee 
Using Eqs.~(\ref{night5}) and (\ref{rho2}), we then obtain 
\be
{\cal S}(E)\simeq S_C(\beta_0)-{1\over 2}\ln (2\pi S''_C(\beta_0))-
\ln(|F'(E)|).
\label{approx}
\ee
This formula that was originally suggested in~\cite{das} had missed the last 
term on the RHS. This has also been pointed out earlier in ref.~\cite{chatter}.   
It turns out that for two of the models that we consider in 
this paper, $F'(E)=1$, and the last term in Eq.~(\ref{approx}) does not 
contribute. On the other hand, in the model with a logarithmic energy 
spectrum, this term plays a crucial role. Note that within the canonical 
formalism,   
$S_C''(\beta_0)=(\langle E^2\rangle -\langle E\rangle^2)$ is the fluctuation 
squared of the energy~\cite{kubo}. When this energy fluctuation is subtracted 
out from the canonical entropy, as in Eq.~(\ref{approx}), we obtain an 
estimate for the microcanonical 
entropy (for quantum gravitational fluctuations, see e.g. \cite{pm2}). 

The approximation (\ref{approx}) for the microcanonical entropy ${\it S(E)}$ 
is very useful, since it is prohibitively difficult to calculate 
it directly from Eq.~(\ref{mike}). Generally, in a mean-field model, one 
is given the single-particle quantum spectrum. The direct computation of the 
many-body degeneracy factor $\Omega(E)$ from this starting point is very 
time consuming. Instead, it is much simpler to obtain the canonical 
$N$-body partition function by well-known recursion relations 
(depending on the 
quantum statistics)~\cite{borrmann}, and then compute the canonical 
entropy $S_c(\beta)$. Going one step further, one may calculate the canonical 
energy fluctuation, and use Eq.~(\ref{approx}) to obtain ${\cal S}$. By 
following this canonical route, no computation of $\Omega(E)$ is necessary. 
The approximate formula (\ref{approx}), relevant to black hole physics,  
has not been yet explicitly tested. The main objective of this 
paper is to test this formula quantitatively in some  
model systems where the microcanonical entropy ${\cal S}$ can be calculated 
exactly.
\subsection{The logarithmic energy spectrum}
In the first idealized example, we consider $N$ noninteracting bosons 
($N\rightarrow \infty$) occupying a set of single-particle energy levels 
(and also the ground state, which is at zero energy) :
\be
\epsilon_p=\ln p~,
\label{primes}
\ee
where $p$ runs over all the prime numbers $2,3,5,...$. As recently pointed 
out in ref.~\cite{prime}, the many-body microcanonical entropy ${\cal S}$ 
of this system is {\it exactly} zero. This is because of the fundamental 
theorem of arithmetic, which states that every positive integer $n$ can be 
expressed only in one way as a product of the prime number powers \cite{euclid}
\be 
n=p_1^{n_1}p_2^{n_2}...p_r^{n_r}...,
\label{funda}
\ee 
where the $p_r$'s are distinct primes, and $n_r$'s are positive 
integers, including zero, and need not be distinct. It immediately follows 
from Eq.~(\ref{funda}) that the eigenenergies of the many-body system 
are given by 
\be
E_n=\ln n=\sum_r n_r\ln p_r~,
\label{log}
\ee 
and that each eigenstate is {\it non-degenerate}. This means that for every 
macro-state of the many-body system, there is exactly one microstate, and 
$\Omega(E_n)=1$.  
This implies, by Eq.~(\ref{mike}), that the microcanonical entropy 
${\cal S}(E_n)=0$. We would now like to check if this is verified from 
Eq.~(\ref{approx}). For the many-body spectrum given by Eq.~(\ref{log}), 
noting that the inverse function $F(E)=\exp(E)$, we immediately obtain 
the density of states 
\be 
\rho(E)=e^E[1+2\sum_{k=1}^{\infty} {\rm cos}(2\pi k e^E)]~.  
\label{exp}
\ee
The second term on the RHS is the intrinsic quantum fluctuation, due to 
$E_n$'s taking only discrete values. The smooth part of the density of 
states is ${\tilde {\rho}}(E)=e^E$. To obtain the canonical entropy from 
Eq.~(\ref{entropy}), we need to calculate the canonical partition function 
$Z(\beta)$. The exact $Z(\beta)$ for $N\rightarrow \infty$ in this case is the 
Riemann zeta function $\zeta (\beta)=\sum_{n=1}^{\infty} n^{-\beta}$, which 
includes the quantum fluctuations. We pick up the smooth part of 
$Z(\beta)$ by evaluating it using Eq.~(\ref{zce}) with 
${\tilde {\rho}}(E)=e^E$, obtaining 
\be 
Z(\beta)=(\beta-1)^{-1},~~~~\beta \geq 1~.
\label{limit}
\ee
From this, we get $S_c(\beta)=-\ln (\beta-1) +\beta E$, so the   
saddle-point is given by $\beta_0=(1/E+1)$. Thus the equilibrium canonical 
entropy is 
\be 
S_c(\beta_0)= E+\ln E+1~,
\ee  
that contains both a linear and a logarithmic term. Evaluation of the 
fluctuation term is elementary, and the microcanonical entropy using 
Eq.~(\ref{approx})  is 
\be
{\cal S}(E)= E+ \ln E +1-{1\over 2} \ln (2\pi E^2) -E =1-{1\over 2}\ln (2\pi).
\label{can}
\ee 
We see that the $E$ dependent terms in the canonical entropy are entirely canceled by the fluctuation 
term, the small residual constant is due to the use of the saddle-point method. 
This example is atypical, because the canonical term contains both the linear 
and the logarithmic terms, and still Eq.~(\ref{approx}) yields (almost) the 
correct microcanonical estimate. 
\subsection{The power-law single-particle spectrum}
For our second example, we consider N noninteracting bosons confined in a  
mean field with a single particle spectrum given by $\epsilon_m =
m^s,$ where the integer $m\geq 0$, and $s>0$. The 
energy is measured in dimensionless units. This model is considered here 
because the canonical partition function (for $N\rightarrow\infty$) is 
exactly known \cite{ramanujan}:
\be
Z(\beta) = \prod_{m=1}^{\infty}\frac{1}{[1-\exp(-\beta m^s)]}~.
\label{zprod}  
\ee
We can therefore calculate the exact microcanonical partition function 
$\Omega (E)$ by expanding this $Z(\beta)$ as in (\ref{zce}).
This is illustrated in the Appendix by taking a quadratic single-particle 
spectrum ($s=2$), and showing that the exact combinatorial result for 
$\Omega(E)$ of the many-body system can be reproduced by expanding the 
canonical partition function above. It also illustrates the important point 
that even though the single-particle energies are not equi-spaced, the 
many-body system has equi-spaced eigenenergies $E_n$'s. This has the 
consequence that $F'(E)=1$, and $\Omega(E)=\tilde{\rho}(E)$ when the quantum 
oscillations are dropped. 
The numerical calculations for a large number of particles, 
using different power-law single-particle spectra, were done in a different 
context in Ref.~\cite{tran}, where the details may 
be found. We test the accuracy of Eq.~(\ref{approx}) by comparing it with  
the exact ${\cal S}(E)$ from Eq.~(\ref{mike}) for $s=1, 2$. The quantum 
oscillations have been included in the exact microcanonical calculations, 
but are difficult to see in this scale. In 
Figs.~1(a) and (b), the dashed curve denotes the canonical entropy $S_C(E)$ 
as given by 
Eq.~(\ref{entropy}), and the continuous curve the exact microcanonical 
entropy ${\cal S}(E)$ for the two power laws. We see from these curves that 
the two differ substantially as a function of the excitation energy $E$, 
specially for $s=2$. Inclusion of the logarithmic 
correction to the canonical entropy using Eq.~(\ref{approx}) results, however, 
in almost perfect agreement, as shown by the dot-dashed curves in these 
figures. 

\section{A Model for an Area Ensemble of a Black Hole} 
The above bosonic model 
with the power law spectrum $\epsilon_n=n^s$ is not directly applicable 
to the black hole problem since $${\tilde \rho}(E) \propto 
E^{-{3s+1\over {2(s+1)}}} \exp\left[\kappa (s+1)E^{{1\over {1+s}}}\right],$$
 where $\kappa$ is independent of $E$~\cite{ramanujan,tran}, 
and the leading term of $\ln {\tilde \rho}(E)$ does not go linearly with $E$ 
(or $A$, when taken over to the area ensemble). 
Following \cite{aps}, we consider instead {\em distinguishable} elementary 
components, and consider the situation where the elementary area components 
are equi-spaced, $a_j=j$ with $j$ taking values $0,1,2,..$ etc. 
For a macroscopic black hole with a horizon area $A$, we assume that the 
number of independent components is 
\be
N=\eta {A\over l_p^2},
\label{free}
\ee
where $\eta$ is a positive constant, and $l_p$ the Planck length, which is set to unity. Both $N$ and $A$ are fixed quantities in the microcanonical picture. 
As shall be shown shortly, the advantage of this model is that the expression 
for the multiplicity $\Omega(A,N)$ may be explicitly found, and therefore 
the exact microcanonical entropy may be calculated directly.  
By expanding the microcanonical entropy for large $A$, we find that the 
leading term is proportional to the area $A$ of the horizon, and the next 
term goes like $\ln A$. This expression is exactly reproduced in the 
canonical ensemble calculation, when the ensemble averaged $\left<A\right>$ is identified with $A$, and the fluctuation in $\left<A\right>$ is subtracted out.    
For simplicity, we take the $a_j$-spectrum to be 
nondegenerate. This is identically the one-dimensional harmonic oscillator 
spectrum in statistical mechanics. The degeneracy $g(j)=(j+1)$, considered 
in \cite{aps}, is equivalent to a two-dimensional harmonic spectrum and 
is also examined. For the first case, the one-body partition function is 
$$Z_1=\sum_j g(j)\exp (-\alpha a_j)=\sum_j \exp (-\alpha j)=
{1\over {1-e^{-\alpha}}}={1\over {1-x}}~,$$
where $\alpha$ is a variable canonical to the area, and $x=exp(-\alpha)$. 
The canonical $N$-particle partition function for 
distinguishable elementary components is 
\bea
Z_N &=& (Z_1)^N~=(1-x)^{-N}, \nonumber \\                      
   &=& 1+Nx+{N(N+1)\over 2} x^2 +{N(N+1)(N+2)\over {3!}} x^3 +..., \nonumber \\
    &=& \sum_{A=1}^{\infty} {\Pi_{i=0}^{A-1} (N+i)\over {A!}} x^{A}, \nonumber \\
    &=& \sum_{A} \Omega^{(1)}(A,N) e^{-\alpha A},
\eea
which is analogous to Eq.~(\ref{zce}).  The multiplicity of states of 
area is therefore 
\be
\Omega^{(1)}(A,N)={\Pi_{i=0}^{A-1} (N+i)\over {A!}}~.
\label{fin}
\ee
This is the microcanonical partition function.
The superscript (1) is to distinguish from the degenerate case, to be 
discussed later. It is not difficult to check by combinatorics that 
$\Omega^{(1)}(A,N)$ for a given $A$ and $N$ is indeed given by 
Eq.~(\ref{fin}).  Note that unlike the case of 
bose statistics where the multiplicity $\Omega(E)$ was found only by 
expanding the partition function (Eq.~(\ref{zce})) or by exact counting, 
here due to the distinguishability property of the system it is given by an 
explicit formula.  Thus an analytical expression for the microcanonical 
${\cal S}(E)$ may be found directly from Eq.~(\ref{fin}):
\bea
{\cal S}(E)      &=& \ln \Omega^{(1)}(A,N), \nonumber \\
                &=& \sum_{i=0}^{A-1} \ln(N+i) -\ln A!, \nonumber \\
                 &\simeq& A\ln(1+{N\over A})+N\ln(1+{A\over N})-1/2 \ln(A+A^2/N)-1/2\ln (2\pi),
\label{fini0}
\eea
where we have used the Stirling's series and the Euler-Maclaurin \cite{stirling} summation formula to approximate the 
sum in step 2 above.

We now calculate the canonical entropy and show that inclusion of the 
logarithmic correction term (Eq.~(\ref{approx})) gives a formula that agrees 
with Eq.~(\ref{fini0}) for the microcanonical entropy.  
The canonical calculations are performed for a fixed $N$, and the ensemble 
averaged area is given by 
\be
\left<A\right>=-{\partial \ln Z_N\over {\partial \alpha}} 
\ee 
at the equilibrium $\alpha_0$. For large $A$, we identify $\left<A\right>=A$, and later 
correct for the fluctuation.
The canonical entropy is 
$$S_C(\alpha)=\alpha A +\ln Z_N=\alpha A-N\ln(1-e^{-\alpha}).$$

The saddle-point is obtained from the condition that $S_C'(\alpha_0)=0$, 
and gives 
\be
\alpha_0=\ln ({N\over A}+1).
\label{bet}
\ee
Therefore,
\bea
S_C''(\alpha_0) &=& {N\over {(e^{\alpha}-1)^2}} e^{\alpha}=A(1+A/N), \\
S_C(\alpha_0) &=& A\ln(1+{N\over A})+N\ln(1+{A\over N}).
\label{can}
\eea
We may now estimate the microcanonical entropy ${\cal S}$ from  
Eq.~(\ref{approx}), where the area fluctuations have been subtracted from  
the canonical $S_c$. In the latter case, we get 
\bea
{\cal S}(A,N)&\simeq& S_C(\alpha_0)-1/2 \ln S_C''(\alpha_0)-1/2 \ln (2\pi)\nonumber,\\
             &\simeq& A\ln(1+{N\over A})+N\ln(1+{A\over N})-1/2 \ln(A+A^2/N)-1/2\ln (2\pi).
\label{fini}
\eea 
This is identical to what we obtained in Eq.~(\ref{fini0}) from 
the asymptotic expansion of the microcanonical entropy. 
Finally, we use the relation (\ref{free}) to eliminate $N$ from the above 
equation. We immediately obtain
\be
{\cal S}(A) \simeq \xi A-1/2\ln A-1/2\ln [2\pi(1+1/\eta)]~,
\label{cool}
\ee 
where $\xi=\ln[(1+\eta)(1+1/\eta)^{\eta}]$. 
Note that the leading term, proportional to $A$, comes from the canonical 
entropy $S_c$, and the correction $\ln A$ arises from the fluctuation in
$\left<A\right>$.
We next consider the second case, where the area spectrum is degenerate.  The degeneracy is given by $g(j)=j+1$.  The same case has been dealt with by the authors in Ref \cite{aps}, but the results were derived in the grand-canonical ensemble.  The one-body partition function in this case is simply
$$Z_1={1\over ({1-x})^2}~.$$
All the calculations including the expression (\ref{cool}) follow as in the nondegenerate case, with $N$ replaced by $2N$.  

We have shown that for a class of statistical mechanical systems with 
power-law single-particle spectrum, the log-corrected entropy formula 
(\ref{approx}) accurately reproduces the microcanonical entropy of these systems. Our results are applicable to a large class of black holes,
with uniformly spaced area spectrum. These include the ones considered in
refs. \cite{bekmukh,equi}. Although in the context of loop quantum gravity, the spectrum found in \cite{rovelli}
is not strictly uniform, it is effectively equi-spaced for large areas. Moreover, as
shown in \cite{aps} and \cite{adg}, exact equi-spaced spectrum may emerge in loop
gravity as well. It would be interesting to extend our analysis to include
charged and rotating black holes, which are described by more than one quantum 
number.

\section{Acknowledgments}
We are grateful to M.V.N. Murthy for some of the results reported in this paper.  We would also like to thank P. Majumdar for his comments and criticism of this 
manuscript. 
This work was supported in part by the Natural Sciences and Engineering 
Research Council of Canada. 

\newpage
\section{appendix}
We give an example of the calculation of the exact entropy for a system of $N$ non-interacting bosons with a single-particle spectrum given by $\epsilon_m=m^2,~m=0,1,2,... $.  Initially, at $T=0$ (or $E=0$), the particles all reside in the ground state, where $m=0$.  Denote by $N_{ex}$ the number of particles in the excited states.  An excitation energy $E$ may be shared by $N_{ex}$ out of $N$ particles such that $E=\sum_i N_i\epsilon_i$, $N_{ex}=\sum_i N_i$.  The multiplicity $\Omega(E,N)$ is the number of ways of doing this.  For example, take $E=8$ and $N=8$, then there are three distinct configurations.  First, $N_{ex}=2$ particles can be excited in which case each takes $2^2=4$ quanta and goes to the second level above the ground state.  In this case, $N_2=2$ and $N_i=0, i \neq 2$. Second, $N_{ex}=5$ particles can be excited, four of which each takes one quantum to the first level ($N_1=4$) above the ground state and the other takes $2^2=4$ quanta to the second level ($N_2=1$).  Finally, all $8$ particles can be excited, each takes one excitation quantum to the first level ($N_1=8$, and $N_i=0, i \neq 1$).  The energy in all three cases are the same:
\bea
E=8= N_1 \epsilon_1+N_2 \epsilon_2 +...&=& 0+ 2\times 2^2+...+0, ~or \nonumber \\
             &=& 4\times 1^2 + 1\times 2^2+...+0, ~or \nonumber \\
             &=& 8\times 1^2+...+0. \nonumber
\eea
Hence, $\Omega(8,8)=3$.  We see that this problem is identical to counting the number of ways that an integer $E$ can be partitioned into sum of squares.  Note that had we taken  $5 \le N < 8$ in this example, then there would only be $2$ configurations instead of $3$, since the last case in which each particle takes one quantum is eliminated.  In number theory this is known as restricted partitioning as opposed to the unrestricted case considered above.  Clearly, as long as $N \ge E$, the number of accessible microstates may be enumerated as if $N=\infty$.  In Table I below we enumerate the multiplicity for several values $E$, assuming unrestricted partitioning. 

Few remarks are in order.  First, as mentioned before, as long as $E \le N$, the enumeration of the multiplicity is $N$-independent. Second, as illustrated above, $E$ is given by a set of consecutive integers, $E_n=n$, even though the single-particle energy spectrum is not equi-spaced.  Each many-body energy level $E_n$ has a degeneracy $\Omega(E_n,N)$.  This is in fact general for any power-law single-particle spectrum and non-interacting particles.  Note that the multiplicity $\Omega(E,N)$ enumerated in the Table is the same as the expansion coefficient of the partition function, i.e.
\bea
Z(x) &=& \prod_{m=1}^{\infty}\frac{1}{[1-x^{m^2}]}, \nonumber \\
     &=& 1+x+x^2+x^3+2x^4+2x^5+2x^6+2x^7+3x^8+4x^9 + ...,\nonumber 
\eea
where $x=e^{-\beta}$, and the power of $x$ corresponds to the many-body energy $E_n$.

\begin{table}[t]
\begin{tabular}{|rl|c|c|c|c|c|}
\hline
\multicolumn{2}{|c|}{$E$}& $~~~N_{ex}~~~$ & $~\omega^{(1)}(E,N_{ex},N)~$&$~\Omega(E,N)~$ \\ \hline\hline
$1$ & $=1^2$             &  $1$  &       $ 1 $       & $1$ \\ \hline
$2$   & $=1^2+1^2$      &  $2$  &       $ 1 $        & $1$ \\ \hline
$3$   &  $~=1^2+1^2+1^2$         &  $3$  &       $ 1 $        & $1$ \\ \hline
$4$   & $=2^2$          &  $1$  &        $ 1 $        &     \\
      & $=1^2+1^2+1^2+1^2$  &  $4$  &       $ 1 $        &  $2$    \\ \hline
$5$   & $=1^2+2^2$      &  $2$  &       $ 1 $        &     \\ 
& $=1^2+1^2+1^2+1^2+1^2$ &  $5$  &       $ 1 $        &  $2$    \\ \hline
$6$    &  $=1^2+1^2+2^2$ &  $3$  &       $ 1 $        &     \\ 
       & $=1^2+...+1^2$ &  $6$  &       $ 1 $        &  $2$    \\ \hline
$7$    & $=1^2+1^2+2^2$ &  $3$  &       $ 1 $        &     \\ 
       & $=1^2+...+1^2$ &  $7$  &       $ 1 $        &  $2$    \\ \hline
$8$    &$=2^2+2^2$       &  $2$  &       $ 1 $        &     \\ 
  & $=1^2+1^2+1^2+1^2+2^2$ &  $5$  &       $ 1 $        &     \\ 
       & $=1^2+...+1^2$ &  $8$  &       $ 1 $        &  $3$    \\ \hline
\end{tabular}
\caption{Calculation of the multiplicity $\Omega(E,N)$ for $N$ bosons at an excitation energy $E$.  The single-particle energy spectrum is given by $\epsilon_m=0,1,4,9,...,m^2$.  For a given integral $E$, the partitioning into the number of $N_{ex}$ excited particles are tabulated in column 1, and the corresponding number $N_{ex}$ in column 2.  The microstates $\omega(E,N_{ex},N)$ are enumerated in column 3.  The last column gives the multiplicity $\Omega(E,N)=\sum_{N_{ex}=1}^{N}\omega(E,N_{ex},N)$.  It is to be noted that for a large excitation energy $E$ (not considered in the table), $\omega(E,N_{ex},N)$ may take on values larger than unity.}  
\end{table}

\begin{figure}[h]
\centering
\caption{1a) Comparison of the exact microcanonical entropy ${\cal S}(E)$ (solid line) and the canonical entropy $S_C(E)$ (dashed line) for $\epsilon=m^s$ spectrum, where $s=1$.  The particles are taken to be $N$ non-interacting bosons, where $N\rightarrow \infty$. The dot-dashed curve, given by Eq.~(\ref{approx}), overlaps with the exact solid curve.  1b) Same as 1a), except $s=2$.}
\end{figure}

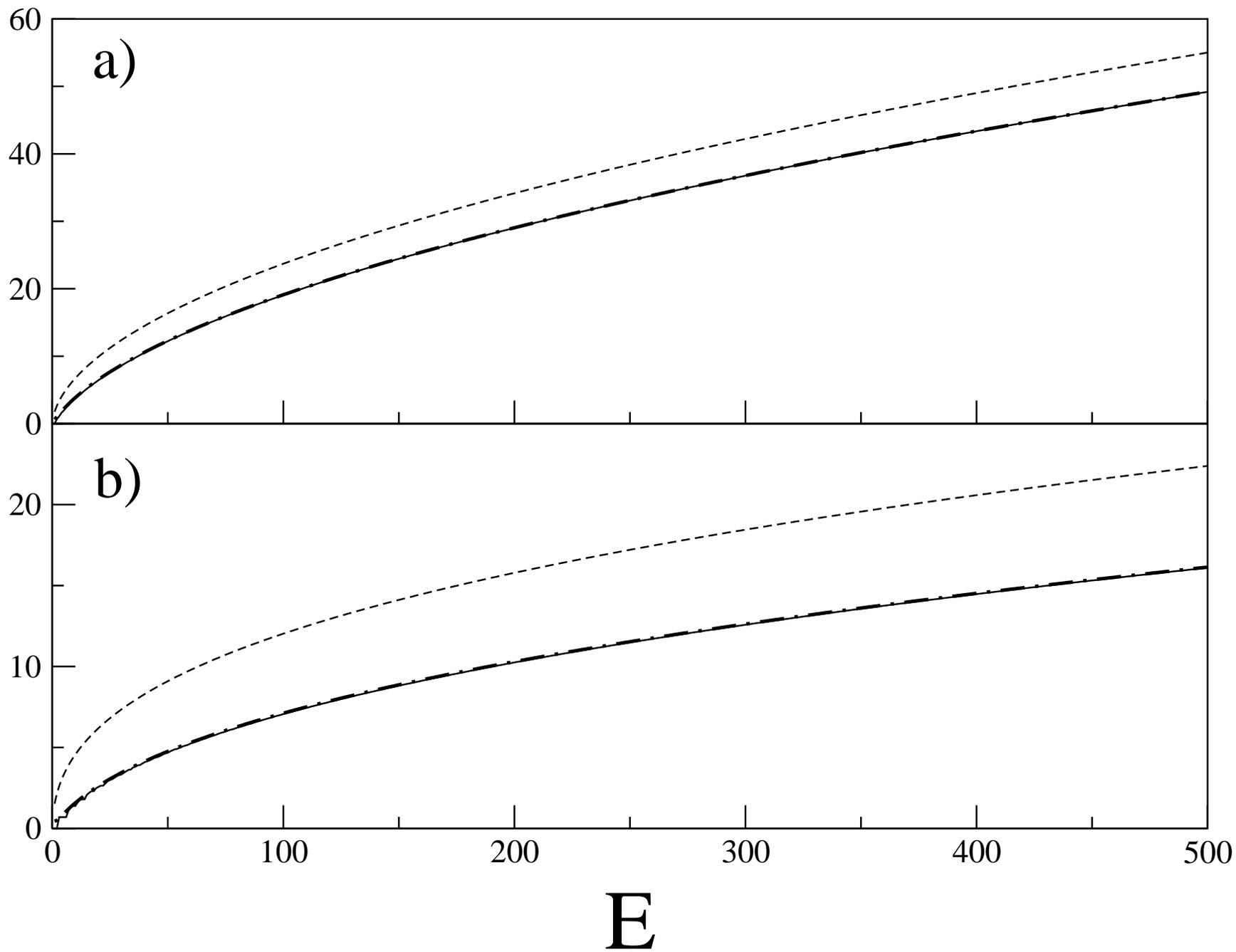
\end{document}